\documentclass[a4paper,nofootinbib,twocolumn]{revtex4}
\usepackage{graphicx}

\begin{document}

\title{Normal phase and superconducting instability in attractive Hubbard
model: the DMFT(NRG) study}
\author{N.A. Kuleeva$^1$\footnote{E-mail: strigina@iep.uran.ru}, E.Z. Kuchinskii$^1$
\footnote{E-mail: kuchinsk@iep.uran.ru}, M.V. Sadovskii$^{1,2}$
\footnote{E-mail: sadovski@iep.uran.ru}}

\affiliation{$^1$Institute for Electrophysics, Russian Academy of Sciences, Ural Branch,
Ekaterinburg 620016, Russia\\
$^2$Institute for Metal Physics, Russian Academy of Sciences, Ural Branch,
Ekaterinburg 620990, Russia}

\begin{abstract}
We study the normal (non-superconducting) phase of attractive Hubbard model
within dynamical mean field theory (DMFT) using numerical renormalization 
group (NRG) as impurity solver. Wide range of attractive potentials $U$ is
considered, from the weak-coupling limit, where superconducting instability is
well described by BCS approximation, up to the strong-coupling region, where
superconducting transition is described by Bose-condensation of compact Cooper
pairs, which are formed at temperatures much exceeding superconducting 
transition temperature. We calculate density of states, spectral density and
optical conductivity in the normal phase for this wide range of $U$, including 
the disorder effects.
Also we present the results on superconducting instability of the normal state
dependence on the attraction strength $U$ and the degree of disorder. Disorder
influence on the critical temperature $T_c$ is rather weak, suggesting in 
fact the validity of Anderson theorem, with the account of the general 
widening of the conduction band due to disorder.
\end{abstract}

\maketitle

\section{Introduction}

The studies of superconductivity in the strong coupling region attracts theorists
for rather long time \cite{Leggett} and most important advance here was made
by Nozieres and Schmitt-Rink \cite{NS}, who proposed an effective approach to
describe crossover from weak coupling BCS limit to the picture of Bose-Einstein 
condensation (BEC) of preformed Cooper pairs in the strong coupling limit.
The recent progress of experimental studies of ultracold gases in magnetic 
and optical traps, as well as in optical lattices, allowed the controlled change
of parameters, such as density and interaction strength 
(see reviews [\onlinecite{BEC1,BEC2}]),  increasing theoretical interests for
studies of superfluidity (superconductivity) for the case of very strong 
pairing interaction, as well as in BCS-BEC crossover region. Probably the 
simplest model allowing theoretical studies of BCS-BEC crossover is the
attractive Hubbard model. It is widely used also for the studies of
superconductor --- insulator transition (see review in  \cite{Trivedi}).
The most effective modern approach to the solution of Hubbard model, both for
strongly correlated electronic systems (SCES) with repulsive interaction and for 
the studies of BCS-BEC crossover in the case of attraction is the dynamical
mean field theory (DMFT), giving an exact solution in the limit of infinite
dimensions [\onlinecite{pruschke,georges96,Vollh10}]. The attractive Hubbard model
was studies within DMFT in a number of recent papers
[\onlinecite{Keller01,Toschi04,Bauer09,Koga11}]. However only few results were
obtained for the normal (non-superconducting) phase of this model, e.g. there
were practically no studies of two-particle properties, such as optical
conductivity.

To describe electronic properties of SCES we obviously need to take into account
different additional interactions (electron-phonon interaction, scattering by
fluctuations of different order-parameters, disorder scattering etc), which are
inevitably present in such systems. Recently we have proposed the generalized
DMFT+$\Sigma$ approach [\onlinecite{JTL05,PRB05,FNT06,UFN12}], which is very
convenient and effective for the studies of such additional (external with respect
to Hubbard model itself) interactions (e.g. pseudogap fluctuations
[\onlinecite{JTL05,PRB05,FNT06,UFN12}], disorder [\onlinecite{HubDis,HubDis2}]
and electron-phonon interaction [\onlinecite{e_ph_DMFT}]). This approach was 
also successfully extended to the analysis of optical conductivity
[\onlinecite{HubDis,PRB07}]. In this paper we apply the DMFT+$\Sigma$ approach
to the studies of the normal state properties of attractive Hubbard model,
including the effects of disorder.

\section{The basics of DMFT+$\Sigma$ approach}

In general case we shall consider non-magnetic Hubbard model with site disorder.
The Hamiltonian of this model can be written as:
\begin{equation}
H=-t\sum_{\langle ij\rangle \sigma }a_{i\sigma }^{\dagger }a_{j\sigma
}+\sum_{i\sigma }\epsilon _{i}n_{i\sigma }+U\sum_{i}n_{i\uparrow
}n_{i\downarrow },  
\label{And_Hubb}
\end{equation}
where  $t>0$ is the transfer integral between nearest sites of the lattice, 
$U$ is the onsite interaction (for the case of attraction $U<0$), $n_{
i\sigma }=a_{i\sigma }^{\dagger }a_{i\sigma }^{{%
\phantom{\dagger}}}$ is onsite electron number operator, $a_{i\sigma }$ 
($a_{i\sigma }^{\dagger}$) is annihilation (creation) operator for electron
with spin $\sigma$ on site $i$, local energy levels $\epsilon _{i}$ 
are assumed to be independent random variables at different lattice sites.
To simplify diagram technique in the following we assume the Gaussian distribution
of these energy levels: 
\begin{equation}
\mathcal{P}(\epsilon _{i})=\frac{1}{\sqrt{2\pi}\Delta}\exp\left(
-\frac{\epsilon_{i}^2}{2\Delta^2}
\right)
\label{Gauss}
\end{equation}
Parameter $\Delta$ represents here the measure of disorder and this Gaussian 
random field (with ``white noise'' correlation on different lattice sites)
generates ``impurity'' scattering and lead to the standard diagram technique for
calculation of the ensemble averaged Green's functions \cite{Diagr}.

Generalized DMFT+$\Sigma$ approach [\onlinecite{JTL05,PRB05,FNT06,UFN12}] 
extends the standard DMFT [\onlinecite{pruschke,georges96,Vollh10}] introducing 
an additional ``external'' self-energy  $\Sigma_{\bf p}(\varepsilon)$ 
(in general case momentum dependent), which is due to some interaction mechanism
outside the DMFT. It gives an effective procedure to calculate both single- and
two-particle properties [\onlinecite{HubDis,PRB07}]. The success of this approach
is connected with the choice of the single-particle Green's function in the
following form:
\begin{equation}
G(\varepsilon,{\bf p})=\frac{1}{\varepsilon+\mu-\varepsilon({\bf p})-\Sigma(\varepsilon)
-\Sigma_{\bf p}(\varepsilon)},
\label{Gk}
\end{equation}
where $\varepsilon({\bf p})$ is the ``bare'' electronic dispersion, 
while the total self-energy neglects the interference between the Hubbard and
``external'' interaction and is given by the additive sum of the local 
self-energy $\Sigma (\varepsilon)$ of DMFT and ``external'' self-energy
$\Sigma_{\bf p}(\varepsilon)$. This conserves the standard structure of
DMFT equations [\onlinecite{pruschke,georges96,Vollh10}]. However, there are
two important differences with standard DMFT. At each iteration of DMFT cycle
we recalculate the ``external'' self-energy $\Sigma_{\bf p}(\varepsilon)$ 
using some approximate scheme for the description of ``external'' interaction
and the local Green's function is ``dressed'' by $\Sigma_{\bf p}(\varepsilon)$ 
at each step of the standard DMFT procedure. 

For ``external'' self-energy due to disorder scattering entering DMFT+$\Sigma$ 
cycle below we use the simplest approximation neglecting the diagrams with
``intersecting'' interaction lines, i.e. the self-consistent Born approximation,
For the Gaussian distribution of site energies it is momentum independent and
is given by:
\begin{equation}
\Sigma_{\bf p}(\varepsilon)\to\tilde\Sigma=\Delta^2\sum_{\bf p}G(\varepsilon,{\bf p}),
\label{BornSigma}
\end{equation}
where $G(\varepsilon,{\bf p})$ is the single-particle Green's function (\ref{Gk}), 
while $\Delta$ is the strength of site energy disorder. 

To solve the single Anderson impurity problem of DMFT we have employed the
reliable algorithm of the numerical renormalization group [\onlinecite{NRGrev}],
i.e. the DMFT(NRG) approach..

Within DMFT+$\Sigma$ approach we can also investigate the two-particle
properties. In particular, the real part of dynamical (optical)conductivity
in  DMFT+$\Sigma$ we have the following general expressionм 
[\onlinecite{HubDis,PRB07}]:
\begin{eqnarray}
{\rm{Re}}\sigma(\omega)=\frac{e^2\omega}{2\pi}
\int_{-\infty}^{\infty}d\varepsilon\left[f(\varepsilon_-)
-f(\varepsilon_+)\right]\times\nonumber\\
\times{\rm{Re}}\left\{\phi^{0RA}_{\varepsilon}(\omega)\left[1-
\frac{\Sigma^R(\varepsilon_+)-\Sigma^A(\varepsilon_-)}{\omega}\right]^2-
\right.\nonumber\\
\left.-\phi^{0RR}_{\varepsilon}(\omega)\left[1-
\frac{\Sigma^R(\varepsilon_+)-\Sigma^R(\varepsilon_-)}{\omega}\right]^2
\right\},
\label{cond_final}
\end{eqnarray}
where $e$ is electronic charge, $f(\varepsilon_{\pm})$ --- Fermi distribution 
for $\varepsilon_{\pm}=\varepsilon\pm\frac{\omega}{2}$, and
\begin{equation}
\phi^{0RR(RA)}_{\varepsilon}(\omega)=\lim_{q\to 0}
\frac{\Phi^{0RR(RA)}_{\varepsilon}(\omega,{\bf q})-
\Phi^{0RR(RA)}_{\varepsilon}(\omega,0)}{q^2}, 
\label{phi0}
\end{equation}
where the two-particle Green's function 
$\Phi^{0RR(RA)}_{\varepsilon}(\omega,{\bf q})$ contain all vertex corrections
from ``external'' interaction, but do not include vertex corrections from
Hubbard interaction. This considerably simplifies calculations of optical
conductivity within  DMFT+$\Sigma$ approximation, as we have only to solve
the single-particle problem determining the local self-energy
$\Sigma(\varepsilon_{\pm})$ via the DMFT+$\Sigma$ procedure. 
Non-trivial contribution from non-local correlations enters only via
$\Phi^{0RR(RA)}_{\varepsilon}(\omega,{\bf q})$, which can be calculated in
appropriate approximation, taking into account only ``external'' interaction.
To obtain the loop contributions 
$\Phi^{0RR(RA)}_{\varepsilon}(\omega,{\bf q})$, determined by disorder
scattering, we can either use the ``ladder'' approximation for the case of
weak disorder, or following Ref. [\onlinecite{HubDis}], we can use the
generalization of the self-consistent theory of localization
[\onlinecite{VW,MS83}], which allows us to treat the case of strong enough 
disorder. In this approach conductivity is determined mainly by the generalized
diffusion coefficient obtained from the generalization of self-consistency
equation [\onlinecite{VW,MS83}] of this theory, which is to be solved in
combination with DMFT+$\Sigma$ procedure. 

In the following we shall consider the three-dimensional system with
 ``bare'' 
semi-elliptic density of states (per elementary cell and one spin projection), 
which is given by:
\begin{equation}
N_0(\varepsilon)=\frac{2}{\pi D^2}\sqrt{D^2-\varepsilon^2}
\label{DOS13}
\end{equation}
with the bandwidth $W=2D$. All calculations below are done for quarter-filled
band ($n$=0.5). The value of conductivity on all figures will be given in 
universal units of $\sigma_0=\frac{e^2}{ha}$ (where $a$ is the lattice spacing).

\section{Main results}

\begin{figure}
\includegraphics[clip=true,width=0.5\textwidth]{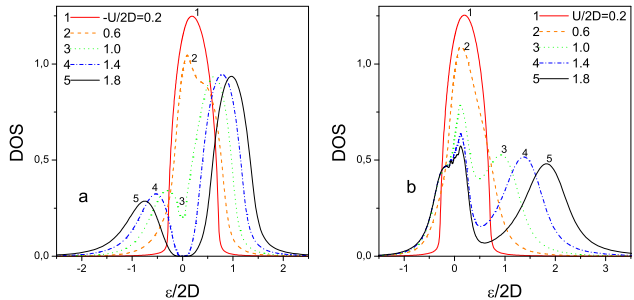}
\caption{Densities of states for different values of Hubbard attraction
(a) and repulsion (b). Temperature $T/2D=0.05$.}
\label{fig1}
\end{figure}

In Fig.\ref{fig1} we show densities of states obtained for $T/2D=0.05$ and quarter filling
of the band ($n=0.5$) for different values of attractive ($U<0$) Fig.\ref{fig1}(a) and 
repulsive ($U>0$) Fig.\ref{fig1}(b) interaction. It is well known that at half-filling 
($n=1$) density of states of attractive and repulsive Hubbard models just coincide (due to 
exact mapping of these models onto each other). This is not so when we deviate from
half-filling. From Fig.\ref{fig1} we can see that the density of states close to the Fermi
level drops with the growth of $U$, both for attraction (Fig.\ref{fig1}(a)) and 
repulsion (Fig.\ref{fig1}(b)), but significant growth of $|U|$ in repulsive case leads only
to vanishing quasiparticle peak and density of states at the Fermi level becomes practically
independent of $U$, while in attractive case the growth of $|U|$ leads to superconducting
pseudogap opening at the Fermi level (curve 3 in Fig.\ref{fig1}(a)) and for $|U|/2D>1.2$ 
we observe the full gap opening at the Fermi level (curves 4, 5 in Fig.\ref{fig1}(a)). 
This gap is not related to the appearance of superconducting state, but is due to the
appearance of preformed Cooper pairs, as the temperature for which the results shown in 
Fig.\ref{fig1} were obtained is larger than superconducting transition temperature 
(cf. Fig.\ref{fig7} below). Thus we observe the important difference from repulsive case, 
where the deviation from half-filling leads to metallic state for arbitrary values of $U$, 
while insulating gap at large $U$ opens not at the Fermi level.

\begin{figure}
\includegraphics[clip=true,width=0.4\textwidth]{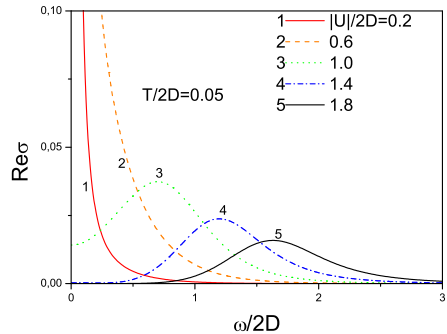}
\caption{Optical conductivity for different values of Hubbard attraction. 
Temperature $T/2D=0.05$.}
\label{fig2}
\end{figure}

This picture of density of states evolution with the growth of $|U|$ is supported by the
behavior of dynamic (optical) conductivity shown in Fig.\ref{fig2}. We see that with the
growth of $|U|$ Drude peak at zero frequency (curves 1, 2 in Fig.\ref{fig2}) is replaced by
pseudogap dip (curve 3 in Fig.\ref{fig2}) and wide maximum of conductivity at finite
frequency, connected with scattering across the pseudogap. The further growth of $|U|$
leads to the appearance of the full gap in optical conductivity due to formation of Cooper 
pairs (curves 4, 5 in Fig.\ref{fig2}).

\begin{figure}
\includegraphics[clip=true,width=0.4\textwidth]{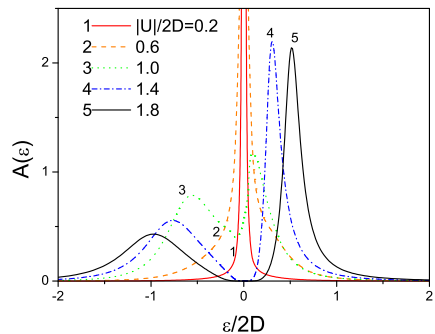}
\caption{Spectral density at the Fermi surface for different values of Hubbard
attraction. Temperature  $T/2D=0.05$.}
\label{fig3}
\end{figure}

\begin{figure}
\includegraphics[clip=true,width=0.4\textwidth]{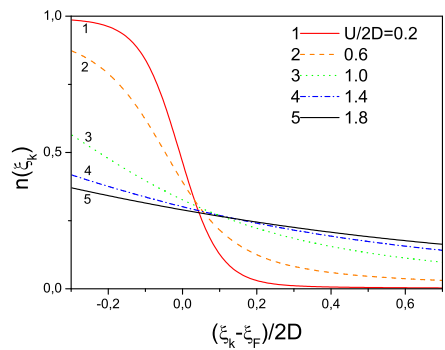}
\caption{Distribution function for different values of Hubbard attraction.
Temperature $T/2D=0.05$. $\xi_F$ -- kinetic energy of electrons at the Fermi surface.}
\label{fig4}
\end{figure}

Similar evolution with growth of $|U|$ is also observed in spectral density.
In Fig.\ref{fig3} we show spectral density 
$A(\varepsilon,{\bf p})=-\frac{1}{\pi}ImG^R(\varepsilon,{\bf p})$ at the
Fermi surface ($p=p_F$) for different values of attractive interaction $U$.
With the growth of $|U|$ a narrow peak in spectral density at the Fermi level 
(curves 1, 2 in Fig.\ref{fig3}) is smeared and with the further growth of $|U|$ 
the pseudogap dip appears at the Fermi level (curve 3 in Fig.\ref{fig3}). 
At still larger $|U|$ this dip is transformed into the real gap (curves 4, 5 in Fig.\ref{fig3}). 
This behavior of spectral density correlates well with qualitative change (with the growth of
$|U|$) of distribution function $n(\xi_k)$ (Fig.\ref{fig4}), defined as:
\begin{equation}
{n(\xi_k)=\int_{-\infty}^{\infty}d\varepsilon A(\varepsilon,{\xi_k})f(\varepsilon)},
\label{nxi}
\end{equation}
where $\xi_k$ represents kinetic energy of electrons. It is seen, that this distribution
changes from more or less defined Fermi step-function at small $|U|$ (curves 1, 2 in Fig.\ref{fig4}) 
to effective constant at large values of $|U|$ (curves 4, 5 in Fig.\ref{fig4}), due to 
formation of Cooper pairs with binding energy of the order of $|U|$.

\begin{figure}
\includegraphics[clip=true,width=0.5\textwidth]{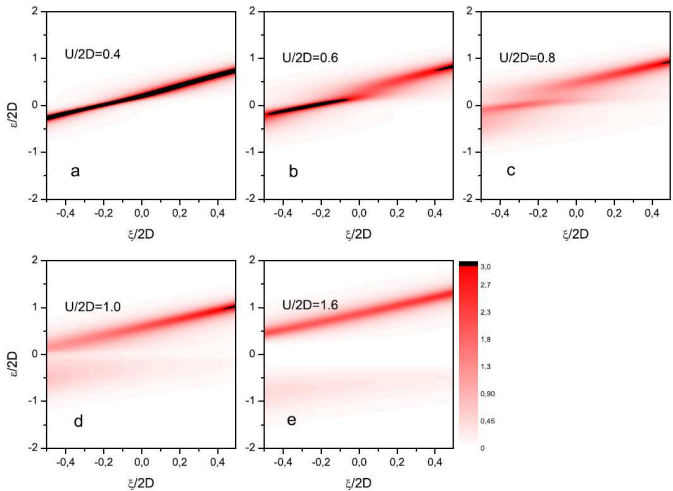}
\caption{Spectral density maps for different values of Hubbard attraction.
Color represent the intensity of spectral density. Temperature $T/2D=0.05$.}
\label{fig5}
\end{figure}

The formation of superconducting pseudogap and Cooper pairing gap with the growth of $|U|$
is also well demonstrated by the maps of spectral density, shown in Fig.\ref{fig5} for 
different values of $U$. Colors represent the intensity of spectral density. We observe that
the growth of $|U|$ leads to transformation of initially well defined dispersion of Fig.\ref{fig5}(a)
to dispersions with pseudogap region, shown in Fig.\ref{fig5}(b,c), which transforms into the
real Cooper gap, shown in Fig.\ref{fig5}(d,e) with the further growth of $|U|$.

\subsection{Disorder effects}

\begin{figure}
\newpage
\includegraphics[clip=true,width=0.48\textwidth]{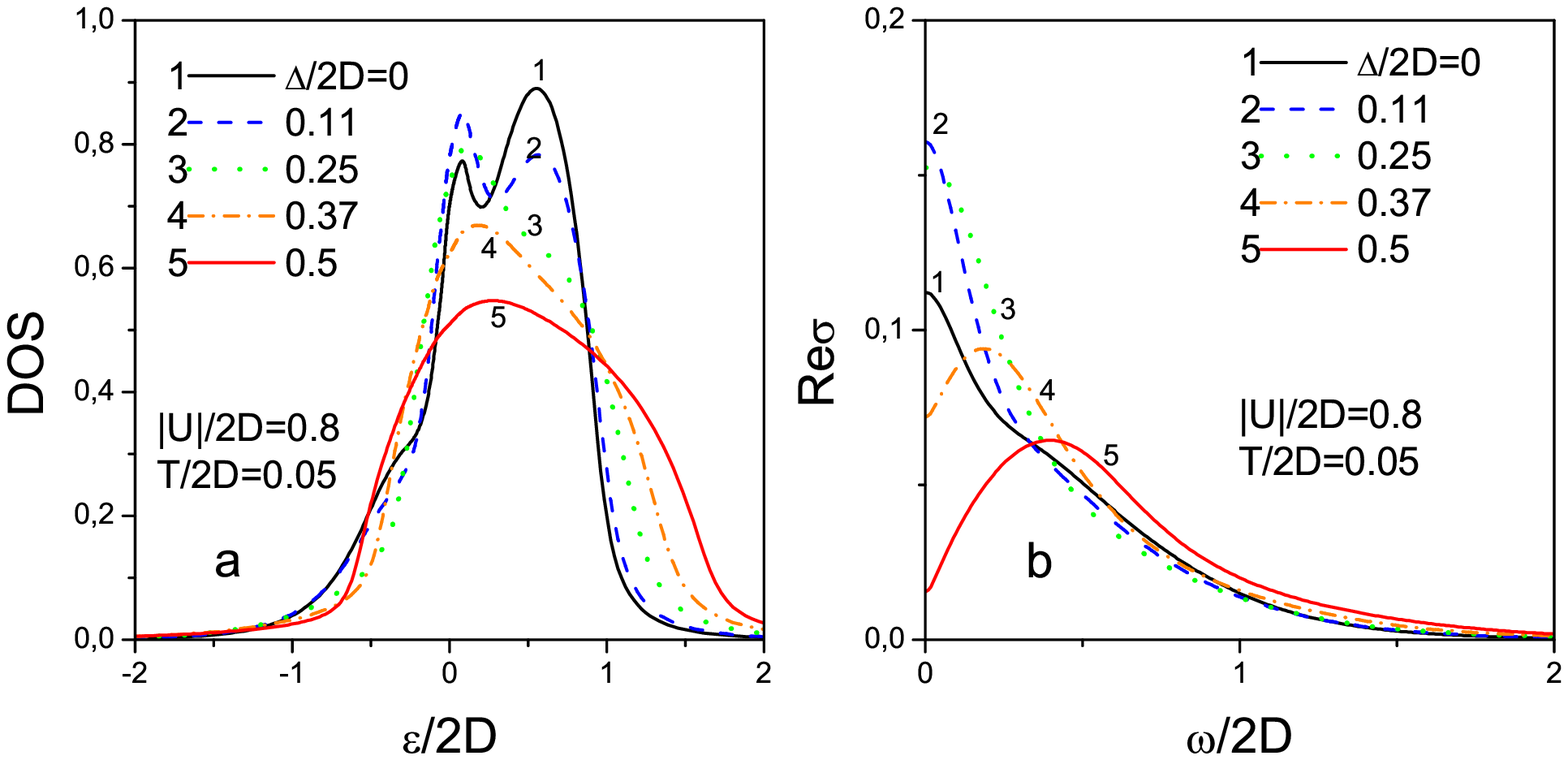}
\includegraphics[clip=true,width=0.48\textwidth]{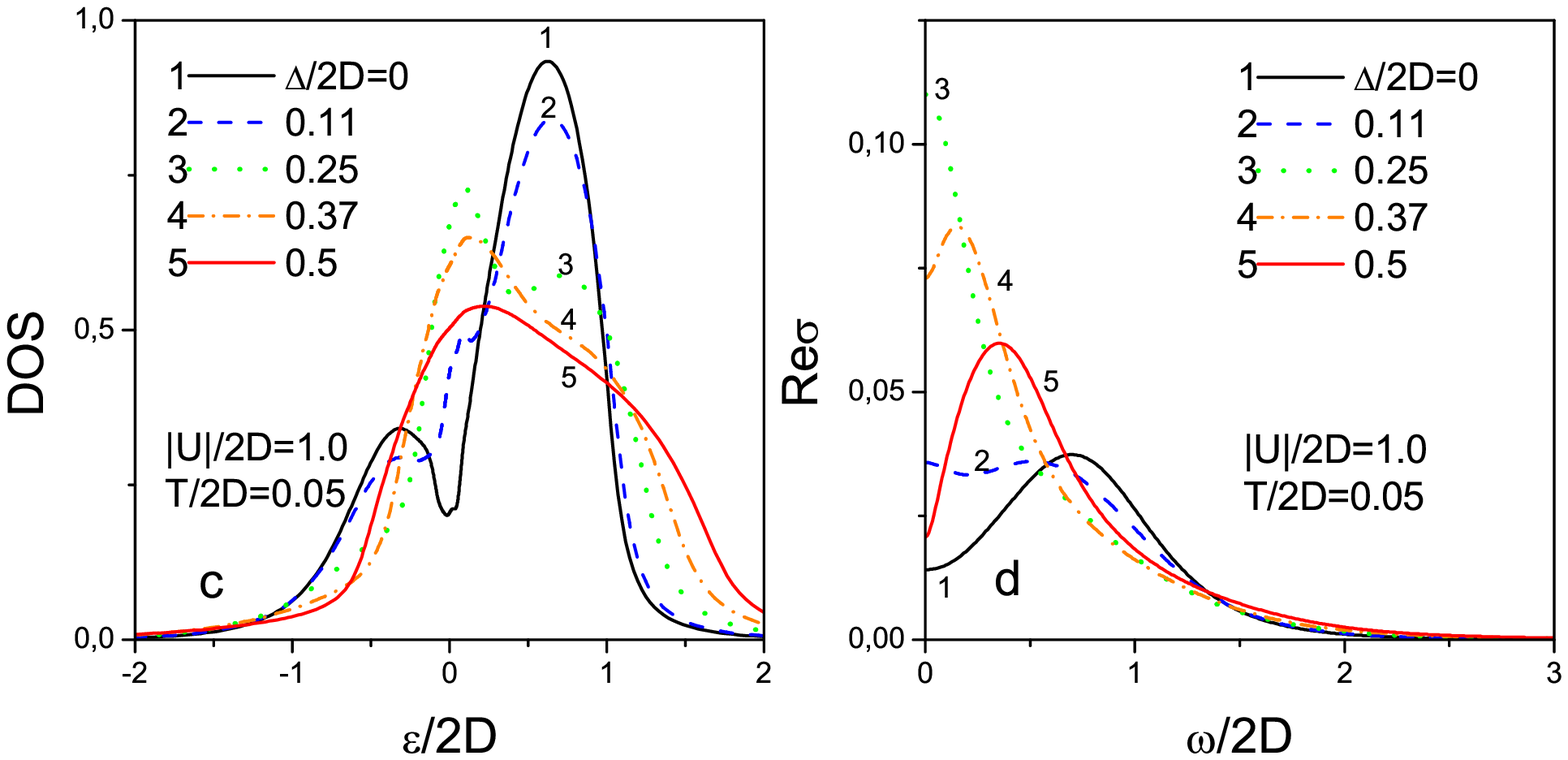}
\includegraphics[clip=true,width=0.48\textwidth]{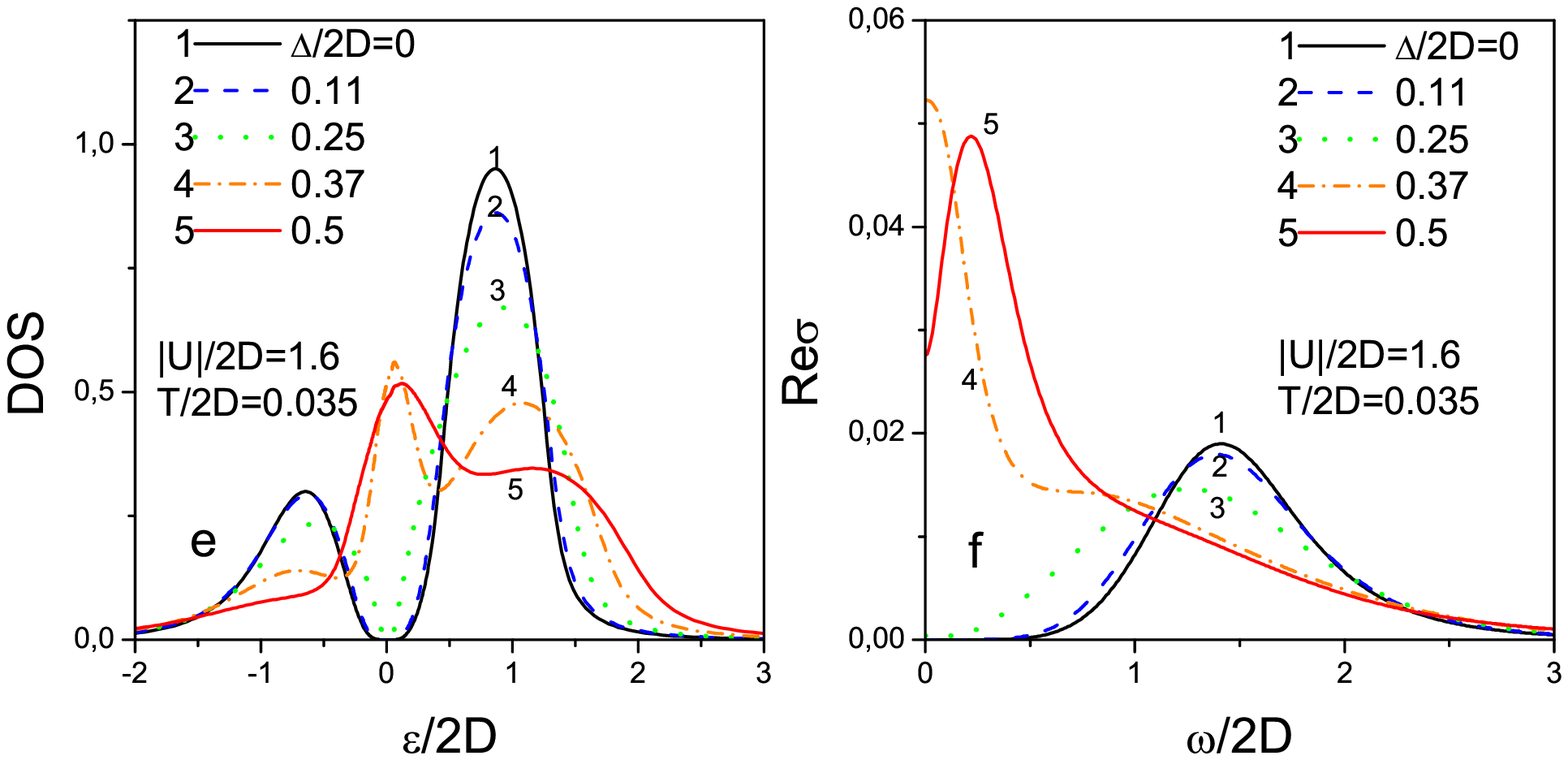}
\caption{Evolution of the density of states (left panels) and optical
conductivity (right panels) with disorder for different values of
$U$ ($|U|/2D=0.8$ - a,b; $|U|/2D=1$ - c,d; $|U|/2D=1.6$ - e,f).}
\label{fig6}
\end{figure}

In Fig.\ref{fig6} we show evolution of the density of states and optical conductivity with
changing disorder. At weak enough attraction ($|U|/2D=0.8$, Fig.\ref{fig6}(a),(b)), we see that
the growth of disorder smears density of states, leading to some widening of the band.
This smearing masks peculiarities of the density of states due to correlation effects.
In particular, quasiparticle peak and ``wings'' due to upper and lower Hubbard bands observed
in the density of states in Fig.\ref{fig6}(a) in the absence of disorder completely vanish at
strong enough disorder. There are no singularities in the density of states due to Anderson
metal-insulator transition, which takes place at $\Delta/2D=0.37$ [\onlinecite{HubDis}], as
density of states does not feel Anderson localization. Evolution of optical conductivity with 
the growth of disorder $\Delta$, shown in Fig.\ref{fig6}(b), corresponds in general to 
evolution of density of states. The growth of disorder, while it remains weak enough,
(curves 1, 2 in Fig.\ref{fig6}(b)), leads to some growth of static conductivity, which 
is connected with suppression of correlation effects at the Fermi level, noted above 
(curves 1, 2 in Fig.\ref{fig6}(a). The further growth of disorder leads to significant
widening of the band and the drop of density of states (curve 3 in Fig.\ref{fig6}(a),(b)), 
which leads to drop of static conductivity. Finally, with the further growth of disorder
Anderson localization effects become important. At $T=0$ Anderson transition takes place at 
$\Delta/2D=0.37$ [\onlinecite{HubDis}]. However, here we consider the case of high enough
temperature $T/2D=0.05$, so that static conductivity (see curves 4, 5 in Fig.\ref{fig6}(b))
remains finite, though at finite frequencies we clearly observe localization behavior 
with $\sigma(\omega)\sim\omega^2$. At larger value of attractive interaction $|U|/2D=1$,
the evolution of the density of states and optical conductivity is more or less similar
(Fig. \ref{fig6}(c,d) ). However, in the absence of disorder we observe here superconducting 
pseudogap in the density of states and disorder growth suppresses it, leading both to the 
growth of the density of states at the Fermi level and appropriate growth of static conductivity.
Finally, at still larger attraction $|U|/2D=1.6$ (Fig.\ref{fig6}(e),(f)) in the absence of 
disorder there is the real Cooper gap in the density of states. This gap is also clearly 
observed in optical conductivity. With the growth of disorder Cooper gap both in the density
of states and conductivity becomes narrower (curves 1-3). Further growth of disorder leads
to complete suppression of Cooper gap and restoration of metallic state with finite density
of states at the Fermi level and finite static conductivity. This closure of Cooper gap is
related to the widening of effective bandwidth $W_{eff}$ due to disorder, which leads to 
the diminishing ratio $|U|/W_{eff}$, which controls the formation of Cooper gap. 
Situation here is similar to the closure of Mott gap by disorder in repulsive Hubbard 
model [\onlinecite{HubDis}].
However, at larger disorder (curve 5 in Fig.\ref{fig6}(f)) we clearly observe localization
behavior, so that the growth of disorder at $T=0$ will first lead to metallic stateе (the
closure of Cooper gap), while the further growth of disorder will induce Anderson 
metal-insulator transition. Similar picture is observed for large positive $U$ at half-filling
($n=1$) [\onlinecite{HubDis}], where the growth of disorder leads to Mott insulator - 
correlated metal - Anderson insulator transition. 

\subsection{Superconducting transition temperature}

\begin{figure}
\includegraphics[clip=true,width=0.4\textwidth]{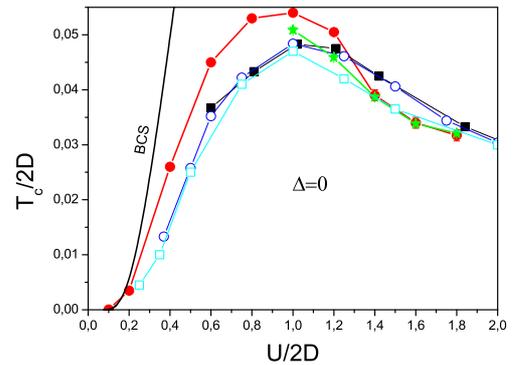}
\caption{Dependence of superconducting critical temperature on attractive
interaction strength. Black squares, white circles and white squares show the
results of Refs. [\onlinecite{Keller01}],[\onlinecite{Toschi04}],[\onlinecite{Koga11}]
respectively for quarter-filled band with $n=0.5$. Stars represent the results
obtained from the criterion of instability of the normal phase.
Filled circles show $T_c$ obtained Nozieres --- Schmitt-Rink approximation.
Continuous black curve represents the result of BCS theory.}
\label{fig7}
\end{figure}

Superconducting transition temperature $T_c$ in attractive Hubbard model was
studied in a number of papers [\onlinecite{Keller01,Toschi04,Koga11}], both 
from the criterion of instability of normal phase (divergence of Cooper
susceptibility) [\onlinecite{Keller01}] and from the condition of vanishing
superconducting order parameter [\onlinecite{Toschi04,Koga11}]. 
In Fig. \ref{fig7} black squares, white circles and white squares show the 
results of Refs. [\onlinecite{Keller01}],[\onlinecite{Toschi04}],[\onlinecite{Koga11}]
correspondingly, for the case of quarter-filling $n=0.5$ (\footnote {In Ref. 
[\onlinecite{Toschi04}] it was claimed that $n=0.75$ was considered, but results
are obtained practically coincide with those of Ref. [\onlinecite{Keller01}] 
obtained for $n=0.5$}). 

Actually, the overall picture of $T_c$ dependence on $U$ is well approximated by
filled circles curve shown in Fig. \ref{fig7} and obtained from Nozieres ---
Schmitt-Rink [\onlinecite{NS}] approach, which gives the correct (approximate)
description of BCS-BEC crossover. 
Then for critical temperature $T_c$ we have the usual BCS-like equation:
\begin{equation}
1=\frac{|U|}{2}\int_{-D}^{D}d\varepsilon N_0(\varepsilon)\frac{th\frac{\varepsilon -\mu}{2T_c}}{\varepsilon -\mu} ,
\label{BCS}
\end{equation}
while the chemical potential for different values of $U$ is to be determined
from DMFT calculations (for fixed band-filling). 
From Fig. \ref{fig7} we can see, that in the weak coupling region of 
$|U|/2D\ll 1$ the critical
temperature in this approach is close to the usual result of BCS theory 
(see appropriate curve in Fig.\ref{fig7}). For $|U|/2D\sim 1$ the critical
temperature $T_c$ has the maximal value, while for $|U|/2D\gg 1$ it drops
as $T_c \sim 1/|U|$ [\onlinecite{NS}], because for such strong values of
attractive interaction the critical temperature is determined by the condition
of Bose condensation of preformed Cooper pairs and transfer amplitude of
these pairs appears only in the second order of perturbation theory and
is proportional to $t^2/|U|$ [\onlinecite{NS}].  
Stars in Fig.\ref{fig7} show the critical temperature, obtained from the
criterion of normal phase instability. For large enough $U$ lowering temperature 
leads to instability of DMFT(NRG) iteration procedure --- at high enough temperatures 
DMFT(NRG) procedure converges to a single solution, while for temperatures below some
critical temperature we observe two different stable solutions for odd or even iterations.
We suggest, that this instability of iteration procedure corresponds to the physical
instability of the normal phase. Unfortunately, for $|U|/2D<1$, the observed instability 
is rather weak (the difference between the odd and even iterations is too small), thus the
accuracy of our calculations is insufficient to determine $T_c$ in this way.
Surprisingly enough, the results for $T_c$ obtained from the approximate 
approach of Ref. [\onlinecite{NS}] and from instability of DMFT(NRG) cycle are 
rather close to each other. This is especially surprising for large values of
$U/2D$ ratio, where pseodigap (or even the real gap) develops in the density of
states.

\begin{figure}
\includegraphics[clip=true,width=0.4\textwidth]{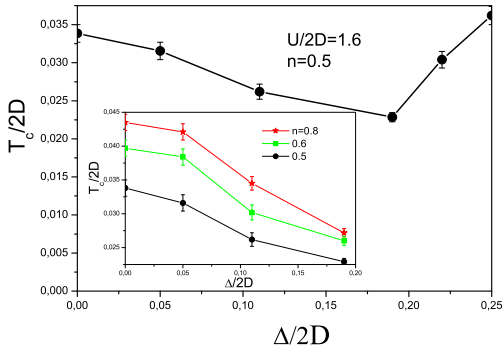}
\caption{Dependence of superconducting critical temperature on disorder for
$|U|/2D=1.6$. At the insert --- $T_c$ suppression by weak disorder for
different values of band-filling: $n=0.5$, $n=0.6$, $n=0.8$.}
\label{fig8}
\end{figure}

In Fig.\ref{fig8} we show
the dependence of critical temperature, obtained from the criterion of normal state instability,
on disorder strength $\Delta$ for $|U|/2D=1.6$. At small $\Delta$ we observe weak suppression 
of $T_c$ by disorder, which is apparetnly due the general smearing of the density of stated and 
bandwidth widening by disorder scattering At large enough.disorder we observe the significant
growth of $T_c$ with the growth of $\Delta$. This is related to the growth of effective
bandwith $W_{eff}$ due to disorder, leading to effective drop of the ration  $|U|/W_{eff}$, 
controlling the value of critical temperature in this model. The growth of disorder leads
to to the drop of  $|U|/W_{eff}$ from the value of  $1.6$ at  $\Delta=0$ to $|U|/W_{eff}\sim 1$ 
for $\Delta/2D\sim 0.4$, which leads to the appropriate growth of the critical temperature
(cf. Fig.\ref{fig7}). This behavior is similar to the growth of the critical value of
repulsion in Hubbard model for Mott metal-insulator transition with the growth of disorder
(cf. Ref. [\onlinecite{HubDis,HubDis2}]). The drop of the ratio $|U|/W_{eff}$ with the growth of
disorder does not allow us to guarantee the sufficient accuracy of the values of 
$T_c$ in the case of $|U|/2D\sim 1$ for disorder values larger than  $\Delta/2D=0.11$. For such  small values of
disorder and for $|U|/2D\sim 1$ the critical temperature is weakly suppressed by disorder, 
similarly to the behavior shown in Fig.\ref{fig8} for the case of $|U|/2D=1.6$.
At the insert in Fig.\ref{fig8} we show the suppression of the critical temperature by weak 
disorder for different values of band-filling: $n=0.5$, $n=0.6$, $n=0.8$.

\section{Conclusions}

Within the generalized DMFT+$\Sigma$ generalization of dynamical mean field
theory we have studied the properties of the normal (non-superconducting) state
of attractive Hubbard model for the wide region of values of onsite attractive 
interaction $U$. The results for the density of states, spectral density,
distribution function and dynamic (optical) conductivity demonstrate the
formation of superconducting pseudogap at the Fermi level for intermediate 
values of coupling strength $|U|/2D\sim 1$ and formation of the real Cooper
gap in the strong coupling region  $|U|/2D>1$. The appearance of  Cooper gap
is related to the formation of compact Cooper pairs at temperatures, which 
are significantly higher. than the critical temperature of superconducting
transition $T_c$, which is determined as Bose-condensation temperature of such
(preformed) pairs. Within our DMFT+$\Sigma$ approach we have also studied the
influence of disorder on the properties of the normal phase. It was shown, that
the growth of disorder in the strong coupling region leads to the closure of
the Cooper gap and restoration of the metallic state, while in the intermediate
coupling region disorder smears superconducting pseudogap and increases the
density of states at the Fermi level. In both cases this is related to the
general widening of the band (in the absence of $U$) by disorder.

We have determined the critical temperature of superconducting transition $T_c$
from the condition of instability of the normal phase. Two methods to find such
instability were used, demonstrating quantitatively similar results. In the weak
coupling region $T_c$ is well described by BCS theory, while in the strong
coupling region it is related to Bose-condensation of (preformed) Cooper pairs
and drops as $1/|U|$ with the growth of $|U|$, passing through the maximum
at $|U|/2D\sim 1$. We have also studied the effects of disorder on $T_c$. 
It was shown, that disorder influence of $T_c$ is rather weak. In the strong 
coupling region, e.g for $U/2D=1.6$ we observe both weak suppression of 
critical temperature, as well as some growth of $T_c$ with the growth of
$\Delta$ for strong enough disorder. In fact, this behavior suggests the
validity of Anderson theorem (as was conjectured for BCS-BEC crossover region
in Ref. [\onlinecite{PosSad}]), with changes of $T_c$ related to the widening of
conduction band by disorder. These results are also consistent with recent
lowest order perturbation theory analysis of the effects of disorder throughout
BCS-BEC crossover region \cite{PalStr}.

This work was partly supported by RFBR grant РФФИ 14-02-00065 and was
performed within the Program of Fundamental Research of the Ural Branch
of the Russian Academy of Sciences ``Quantum macrophysics and nonlinear 
dynamics''(projects No. 12-$\Pi$-2-1002, 12-T-2-1001).

\end{document}